\documentclass[twocolumn,titlepage,preprintnumbers,amsmath,amssymb,article.cls,superscriptaddress]{revtex4-1}
\usepackage{color}
\usepackage{graphicx}
\usepackage{textcomp} 
\usepackage[caption=false]{subfig} 
\usepackage{enumerate}
\usepackage{paralist}
\usepackage{ulem}
\usepackage{eqnarray}
\usepackage{subeqnarray}
\begin{document}
\title{Influence of Slip on the Plateau-Rayleigh Instability on a Fibre}
\author{Sabrina Haefner}
\affiliation{Department of Experimental Physics, Saarland University, D-66041 Saarbr\"ucken, Germany}
\affiliation{Department of Physics \& Astronomy, McMaster University, Hamilton, Canada}
\author{Michael Benzaquen}
\affiliation{PCT Lab, UMR CNRS 7083 Gulliver, ESPCI ParisTech, PSL Research University, Paris, France}
\author{Oliver B\"aumchen}
\affiliation{Department of Physics \& Astronomy, McMaster University, Hamilton, Canada}
\affiliation{Max Planck Institute for Dynamics and Self-Organization (MPIDS), 37077 G\"ottingen, Germany}
\author{Thomas Salez}
\affiliation{PCT Lab, UMR CNRS 7083 Gulliver, ESPCI ParisTech, PSL Research University, Paris, France}
\author{Robert Peters}
\affiliation{Department of Physics \& Astronomy, McMaster University, Hamilton, Canada}
\author{Joshua D.\ McGraw}
\affiliation{Department of Experimental Physics, Saarland University, D-66041 Saarbr\"ucken, Germany}
\author{Karin Jacobs}
\affiliation{Department of Experimental Physics, Saarland University, D-66041 Saarbr\"ucken, Germany}
\author{Elie Rapha\"{e}l}
\affiliation{PCT Lab, UMR CNRS 7083 Gulliver, ESPCI ParisTech, PSL Research University, Paris, France}
\author{Kari Dalnoki-Veress}\email{dalnoki@mcmaster.ca}
\affiliation{Department of Physics \& Astronomy, McMaster University, Hamilton, Canada}
\affiliation{PCT Lab, UMR CNRS 7083 Gulliver, ESPCI ParisTech, PSL Research University, Paris, France}

\begin{abstract}
The Plateau-Rayleigh instability of a liquid column underlies a variety of fascinating phenomena that can be observed in everyday life. In contrast to the case of a free liquid cylinder, describing the evolution of a liquid layer on a solid fibre requires consideration of the solid-liquid interface. In this article, we revisit the Plateau-Rayleigh Instability of a liquid coating a fibre by varying the hydrodynamic boundary condition at the fibre-liquid interface, from no-slip to slip. While the wavelength is not sensitive to the solid-liquid interface, we find that the growth rate of the undulations strongly depends on the hydrodynamic boundary condition. The experiments are in excellent agreement with a new thin film theory incorporating slip, thus providing an original, quantitative and robust tool to measure slip lengths. \end{abstract}

\maketitle

Glistening pearls of water on a spider's web~\cite{zheng}, or the breakup of a cylindrical jet of water into droplets, are familiar manifestations of the Plateau-Rayleigh instability (PRI)~\cite{Plateau:1873wc,Rayleigh:1878vn}. By evolving into droplets, the surface area of the liquid, and consequently the surface energy, are reduced. This instability also acts for a liquid film coating a solid fibre~\cite{Goren:1964uz, Donnelly:1966fo, ROE:1975tw, Eggers:1997uj,  Kliakhandler:2001hp, Craster:2006cu, Craster:2009ca, Duprat:2009kr, Gonzalez:2010iw}, though the flow boundary condition at the solid-liquid interface provides additional complexity to the system. While the breakup of a homogeneous film into droplets on a fibre may be a nuisance in coating technologies of e.g. wires and optical fibres, this fundamental instability turns out to be very useful: take for example water collection through fog harvesting~\cite{Bai:2011gx, Ju:2012gl}, a biomimetic approach that is perfected in nature by the spider's web~\cite{zheng}.  Hence, taking advantage of the PRI on a fibre, enables gaining physical insight into the solid-liquid boundary condition. Indeed, understanding the break-down of the no-slip boundary condition~\cite{Lauga} is of major interest to the scientific and industrial communities as it has practical implications in areas that involve small-scale fluid systems, such as lab-on-a-chip devices, flows in porous media as well as biological flows to name a few.

The classical case of the breakup of a laminar free flowing liquid stream is dependent on material properties and geometry~\cite{EGGERS:2008tw}. While there are some studies of this instability in geometries other than free cylindrical liquid flows~\cite{BrochardWyart:1992kq, Sharma:1996tg, Khare:2007jn, Pairam:2009ud, Villermaux:2009cc, McGraw:2010fn, Shabahang:2011ct}, the role of the boundary condition at the solid-liquid interface on the evolution of the PRI is less well understood~\cite{Choi:2006fa, Muench:2011cu}. It has been shown that the appearance of a dewetting liquid rim undergoing a Plateau-Rayleigh-type instability is influenced by the hydrodynamic boundary condition at the solid-liquid interface~\cite{Baumchen:U9oSTUAs}; however, a general quantitative match between the growth dynamics of the PRI in experiments and analytical theory involving hydrodynamic slip has thus far not been achieved.
\begin{figure}  
\includegraphics[width=1.0\columnwidth]{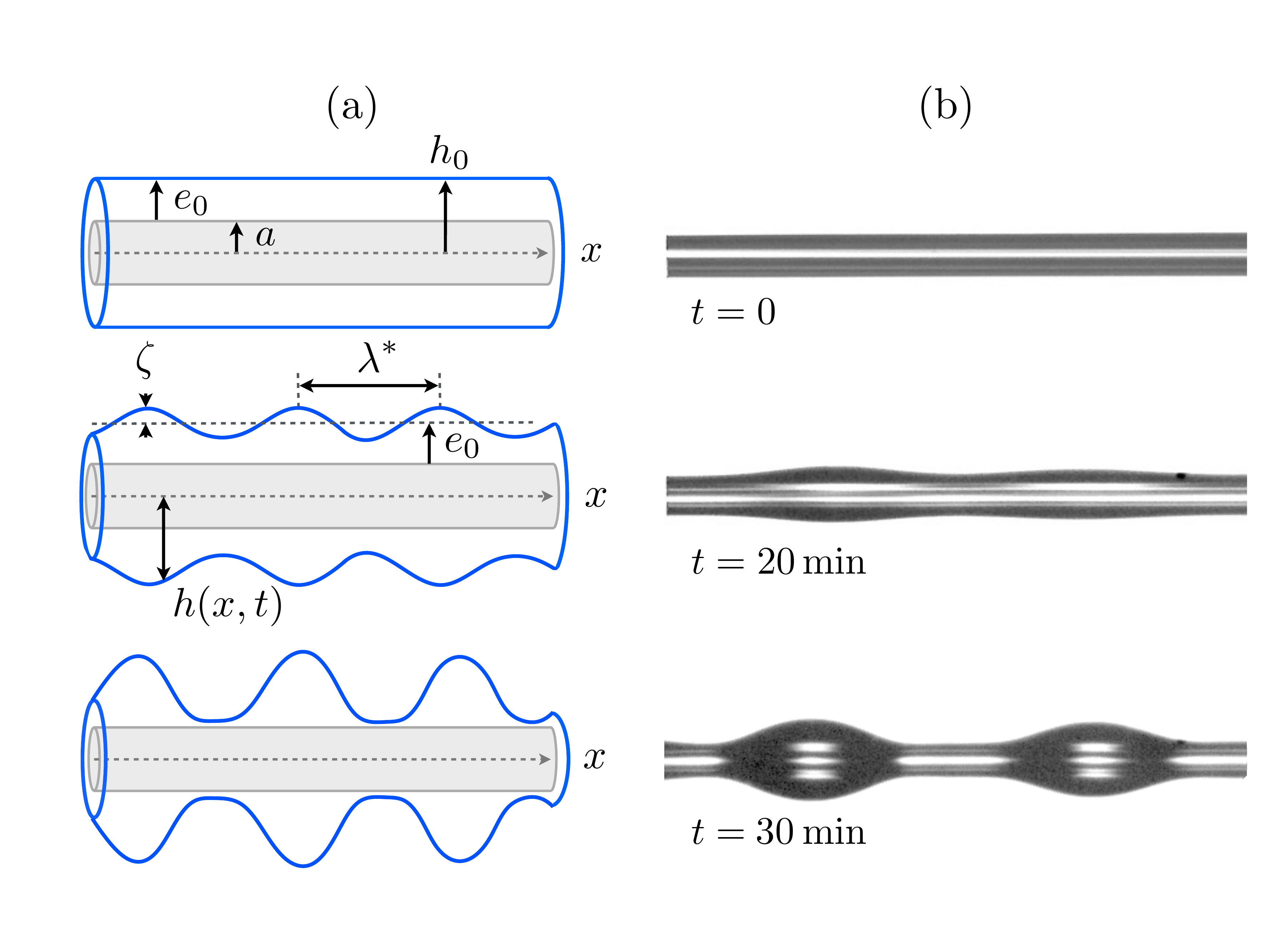}  
\caption{\textbf{Plateau-Rayleigh instability on a fibre.} (a)~Schematic and (b) optical micrographs illustrating the PRI for a liquid polystyrene film on a glass fibre. At $t=0$, the polystyrene film on the no-slip fibre has a thickness $e_0=13.2 \pm 1\,\mu$m and the glass fibre radius is $a=9.6 \pm 1\,\mu$m. The width of the optical images is $560\,\mu$m.}
\label{schematic}
\end{figure} 

Characterisation of slip boundary conditions at interfaces and their controlling parameters has been actively investigated in the literature~\cite{BROCHARD:1992vy, Neto, Schmatko:2005gh, Bocquet:2007kn, Fetzer:2007kt}. The classical boundary condition assumes no slip at the solid-liquid interface. That is, for simple shear flow in the $x$ direction along a solid interface placed at $z=0$, the tangential flow velocity, $v_x(z)$, vanishes at the solid-liquid interface: $v_x(0)=0$. However, as already noted by Navier~\cite{Navier:1823tu}, there is no fundamental principle requiring $v_x(0)=0$, and one can also have hydrodynamic slippage at this interface, as defined by the  slip length: $b=[v_x/\partial_z v_x]|_{z=0}$, where $b=0$ corresponds to the classical no-slip case. 

In this article, we explicitly address the effect of a varying boundary condition, from no-slip to slip at the solid-liquid interface, on the PRI of a viscous liquid layer on a solid fibre. We find that the growth rate of the instability is strongly affected by the solid-liquid interface, with a faster breakup into droplets for a slip boundary condition compared to an equivalent sample with no-slip. In contrast, the wavelength $\lambda^*$ of the fastest growing mode is not sensitive to the solid-liquid interface. The linear stability analysis of a newly developed thin film equation incorporating slip is in excellent agreement with our data. The theory is valid for all Newtonian liquids and enables the precise determination of the capillary velocity $\gamma/ \eta$, and most importantly of the slip length $b$.

\section*{Results}
\textbf{The experimental approach.} An entangled polystyrene (PS, $78\,\mathrm{kg/mol}$) film with homogeneous thickness $e_0$ (5 to 93 $\mu$m), is coated onto a fibre with radius $a$ (10 to 25 $\mu$m), resulting in a PS coated fibre with radius $h_0=a+e_0$, as schematically shown in Fig.~\ref{schematic}(a). Glass fibres provide a simple no-slip boundary condition \cite{McGraw:2013to}. In contrast, a slip interface results from coating the entangled PS film onto a glass fibre pre-coated with a nanometric thin amorphous fluoropolymer (AF2400, 14 $\pm$ 1 nm)~\cite{Baumchen:2009in}. The fluoropolymer coating on glass was used because it is well established that PS, above a critical molecular weight ($M_{\textrm{c}} \sim 35\,{\mathrm{kg}}/{\mathrm{mol}}$), exhibits significant hydrodynamic slip at this solid-liquid interface~\cite{Baumchen:2009in}. Henceforth, we will refer to these as the ``no-slip"  and ``slip" fibres.

All samples were prepared and stored at room temperature, well below the PS glass transition temperature ($T_{\textrm{g}} \sim 100\,^\circ \mathrm{C}$), thereby ensuring that there is no flow in the PS film prior to the start of the experiment. Prior to each experiment, the PS coated fibre was measured with optical microscopy as shown in the $t=0$ image of Fig.~\ref{schematic}(b). To initiate the experiments, samples were annealed in ambient atmosphere at $180\,^\circ$C -- well above $T_{\textrm{g}}$ -- which causes the PRI to develop. The evolution of the surface profile was recorded with optical microscopy. Figure~\ref{schematic}(b) shows a typical evolution for a PS film on a no-slip fibre. 

Above $T_{\textrm{g}}$, the liquid PS film becomes unstable, which causes variations in the local axisymmetric surface profile, $h(x,t)=h_0+\zeta(x,t)$, over the axial coordinate $x$ and time $t$. The amplitude $\zeta(x,t)$ of the undulations grows with time and finally results in a droplet pattern displaying a uniform wavelength, $\lambda^*$. By measuring the spatial variation of $\zeta(x,t)$ in the initial development, and locating the maxima,  the PRI wavelength of each sample was determined (see Fig.~\ref{schematic}(a)). Typically, four or five wavelengths were averaged per sample. In addition, by measuring the temporal change in radius of an individual bulge, we gained information about the growth rate of the instability, and consequently the influence of the hydrodynamic boundary condition. 
\bigskip

\textbf{The theoretical approach.} The experimental findings can be understood within the lubrication approximation, from a thin-film model based upon the Laplace pressure-driven Stokes equation. We assume incompressible flow of a viscous Newtonian liquid film, of thickness varying from 5 to 93 um which is well above the film thickness where disjoining pressure plays a role (a few tens of nanometers) \cite{Seemann:2001iq}. Gravitational effects can be neglected since all length scales involved in the problem are well below the capillary length $l_{\protect \textnormal {c}}\simeq1.73\protect \tmspace +\thinmuskip{.1667em}$mm. Finally, the velocities of the liquid films are small (e.g. the fastest observed rate of change in the amplitude is $\sim25\protect \tmspace +\thinmuskip {.1667em}$nm/s). Thus, the Reynolds and Weissenberg numbers are orders of magnitude smaller than 1 and inertial and viscoelastic effects can be ignored.

We non-dimensionalise the problem (see Fig.~\ref{schematic} for variable definitions) through:
\begin{equation}
H_0=\frac{h_0}{a}, \, H=\frac{h}{a}, \,  X=\frac{x}{a}, \, \Lambda^*=\frac{\lambda^*}{a},  \,  B=\frac{b}{a}, \,  T= \frac{\gamma}{\eta}\frac{t}{a},
\label{rescaling}
\end{equation}
where the capillary velocity $\gamma/\eta $ is the ratio of the liquid-air surface tension to the viscosity of PS. By assuming volume conservation, no stress at the liquid-air interface, and the Navier slip condition at the solid-liquid boundary, one obtains (see Supplementary Methods) the governing equation for the dimensionless profile $H(X,T)$:
\begin{eqnarray}\label{FTFEADIM}
  \partial_TH + \frac{1}{H} \left[\frac{H'+H^2H'''}{16}  \mathcal{M}(H,B)\right]'&=&0 \ ,
\end{eqnarray}
where: $\mathcal{M}(H,B) =$
\begin{equation}
4H^2 \log H + (4B-3)H^2+ 4-8B+\frac{4B-1}{H^2} \ ,  \label{aux}
\end{equation}
and where the prime denotes the partial derivative with respect to $X$. Note that Eq.~\eqref{FTFEADIM} is a composite equation in the sense that we have kept a second order lubrication term in the pressure contribution: the axial curvature. It is the lowest order term counterbalancing the driving radial curvature, and it is thus crucial to obtain the actual threshold of the instability. The dynamical aspect of the flow is however well described at the lowest lubrication order. An interesting discussion on this matter can be found in Ref.~\cite{Craster:2006cu}.

Performing linear stability analysis, namely letting $H(X,T)=H_0+\varepsilon (T)e^{iQX}$, where $\varepsilon (T)\ll 1$, yields an exponential growth of the perturbation of the form $\varepsilon (T)\propto e^{T/\tau(Q)}$, where the rate function is given by:
\begin{equation}
\frac{1}{\tau(Q)} =Q^2\left(1-H_0^{\,2}Q^2\right)\frac{\mathcal{M}(H_0,B)}{16H_0}\ .
\label{Ratefun}
\end{equation}
We define the fastest growing mode $Q^*$ corresponding to the smallest time constant $\tau^*=\tau(Q^*)$, and obtain
$Q^{*}={1}/{ ( H_0\sqrt{2}})$.
Writing $Q^*$ in terms of the dimensionless wavelength leads to:
 \begin{equation}
\Lambda^{*}= 2\pi \sqrt{2}  H_0 \ ,
\label{Lambda}
\end{equation}
which is  similar to the classical PRI dominant wavelength for inviscid jets with no solid core. We note that the correspondence between Rayleigh's result and ours is dependent on both the assumption of Stokes flow and on the presence of a solid core in our system.

The dimensionless growth rate of the fastest growing mode is given by:
\begin{equation}
\frac{1}{\tau^*} = \alpha B + \beta \ ,
\label{growth rate}
\end{equation}
where $B$ is the dimensionless slip length, defined in Eq.~\eqref{rescaling}, and where $\alpha$ and $\beta$ are parameters that depend on geometry only:
\begin{subeqnarray} \label{growth rate aux}
\alpha &=& \frac{1}{16H_0^{\,3}} \left(H_0-\frac{1}{H_0}\right)^2\\
\beta &=& \frac{1}{64 H_0^{\,3}} \left(4H_0^{\,2} \log H_0 -3 H_0^{\,2} - \frac{1}{H_0^{\,2}}+4 \right) \ .
\end{subeqnarray}
As one sees, the wavelength of the fastest growing mode depends exclusively on the initial total radius, while the corresponding growth rate is a linear function of the slip length. Note that in the case of shear-thickening liquids, the PRI should be significantly slowed down as a result of the increasing viscosity associated with an increasing strain rate. Conversely, a shear-thinning liquid is expected to accelerate the rise of the instability. In the case of a viscoelastic material, Maxwell-like rheological models \cite{Benzaquen:2014cf} can be implemented and may reveal interesting physics, beyond the scope of the present study.
\begin{figure}[t!]
\begin{center}     
\includegraphics[width=1\columnwidth]{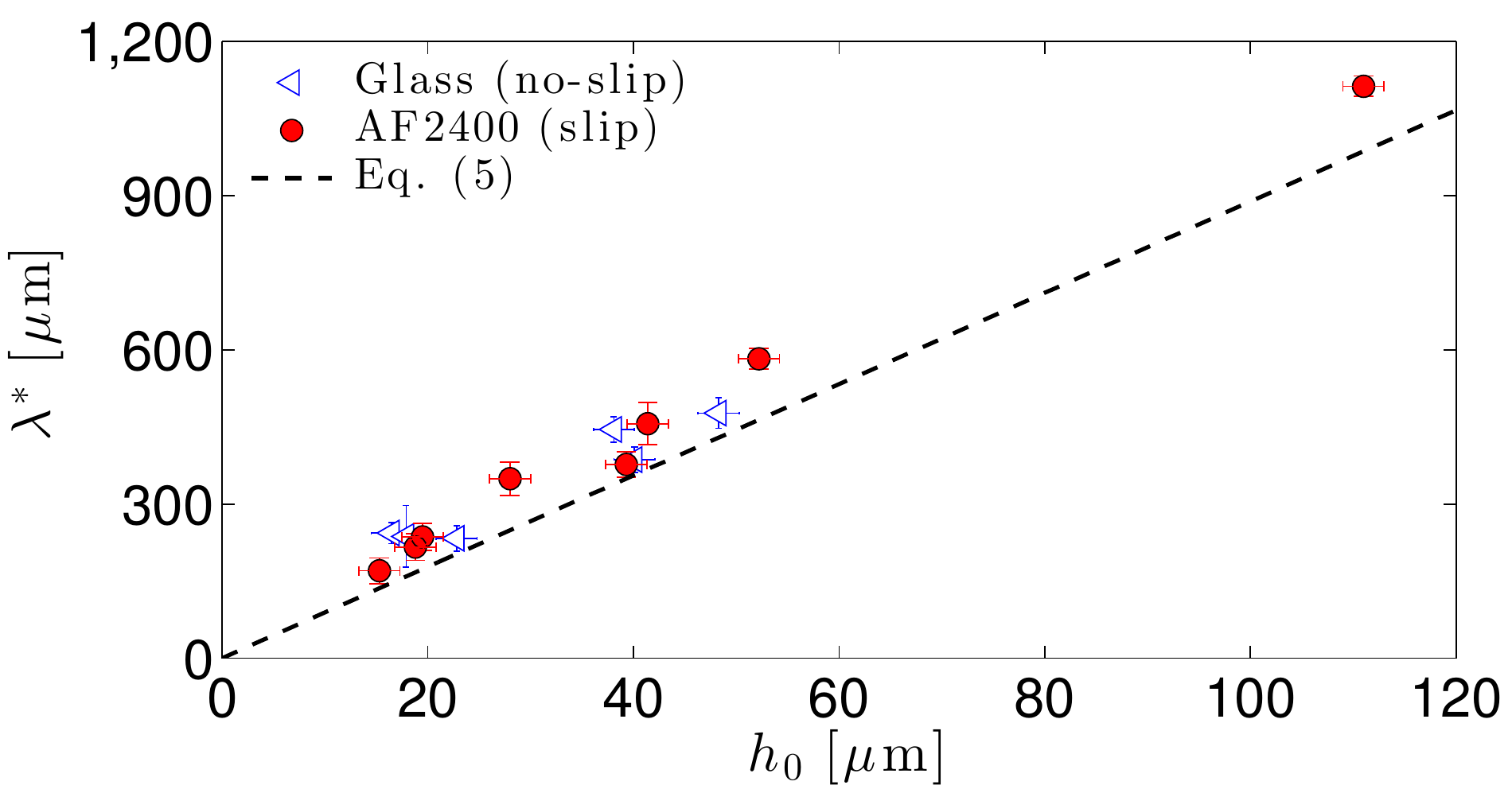}  
\end{center}
\caption{\textbf{Influence of the geometry on the wavelength of the instability.} Wavelength  of the fastest growing mode as a function of the total initial radius  (see Fig.~\ref{schematic}~(a)). The black dashed line represents Eq.~(\ref{Lambda}). The error bars are calculated from the error in the geometry and the inaccuracy given by the wavelength measurement.}
\label{wavelength}
\end{figure} 
\bigskip

\textbf{The spatial evolution of the instability.} Figure~\ref{wavelength} displays the wavelength $\lambda^*$ of the fastest growing modes, measured on no-slip  and slip fibres, as a function of the initial total radius $h_0$. As expected from Eq.~\eqref{Lambda}, the wavelength $\lambda^*$ grows linearly with increasing radius of the fibre-polymer system, and is identical on slip and no-slip fibres. This is consistent with experiments and a theoretical framework for retracting liquid ridges on planar substrates~\cite{Baumchen:U9oSTUAs}. The spatial morphology of the instability at short times is thus unaffected by the solid-liquid boundary condition.
While it is clear from Fig.~\ref{wavelength}  that the wavelength is the same on slip and no-slip fibres, there is a small systematic deviation from the theory. This slight deviation could perhaps be related to the lowest lubrication order of the present model~\cite{Craster:2006cu}, but could also be attributed to the experimental contribution of several modes and the asymmetry of the rate function (see Eq.~\eqref{Ratefun}) in the vicinity of the fastest growing mode.
\begin{figure}[t]
\begin{center}     
\includegraphics[width=1\columnwidth]{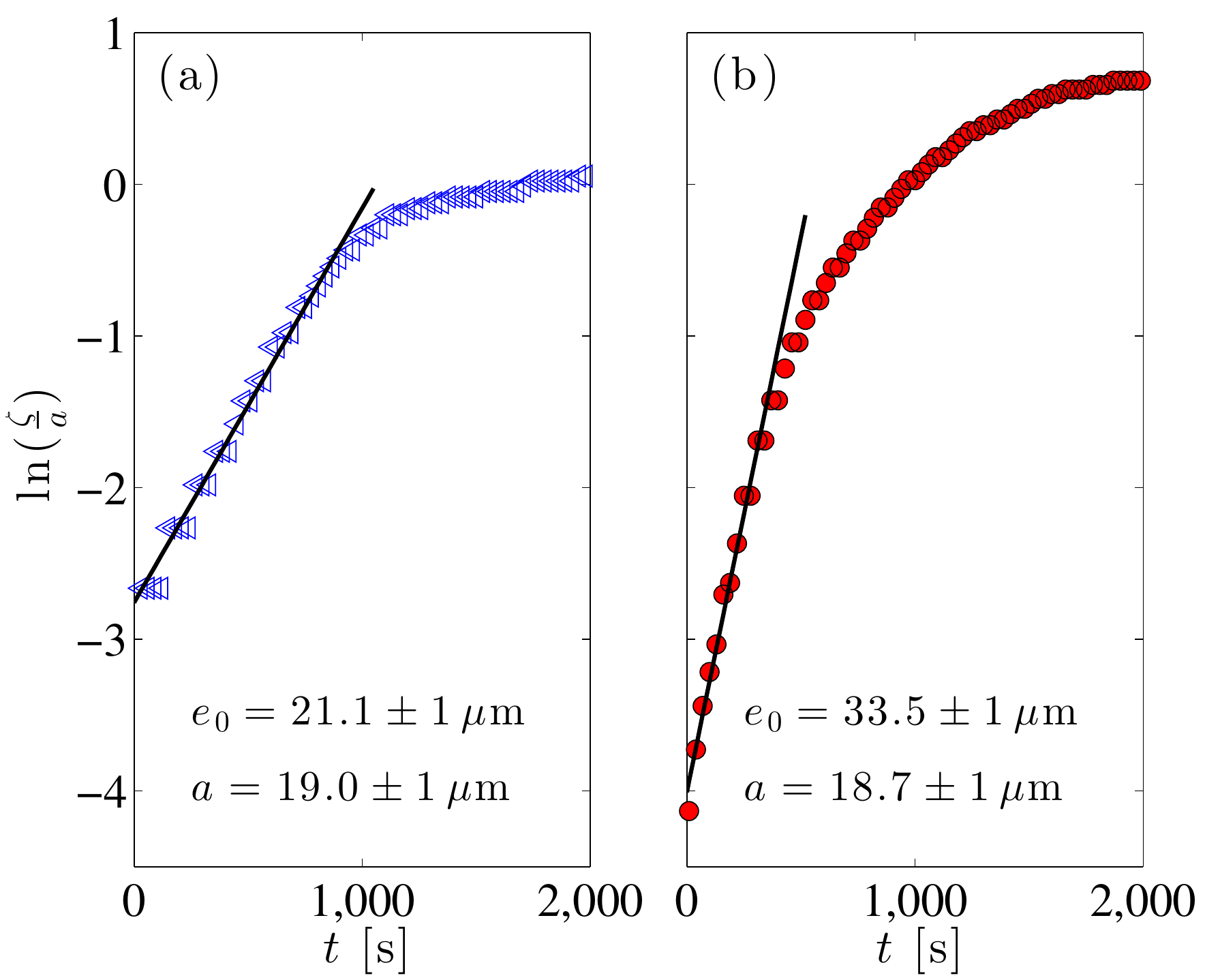}
\end{center}
\caption{\textbf{Temporal growth of the perturbation.} Semi-logarithmic plot of the evolution of the perturbation on (a) a no-slip fibre (glass), and (b) a slip fibre (glass coated with AF2400). The radius of the fibre $a$ and initial thickness of the polymer film $e_0$ (see Fig.~\ref{schematic}~(a)) are indicated. The solid line is the best linear fit in the initial regime.}
\label{fits}
\end{figure}  
\bigskip

\textbf{The temporal evolution of the instability.} We now turn from the spatial morphology of the instability to the temporal evolution. From the experimental images (see Fig.~\ref{schematic}~(b)), we extract the maximal radius of an individual bulge as it develops, in order to obtain the amplitude $\zeta$ as a function of time. The linear stability analysis presented predicts a perturbation that grows exponentially with a dimensionless growth rate $1/\tau^*$ for the fastest growing mode (see Eq.~(\ref{growth rate})). Figure~\ref{fits} displays typical data for the logarithm of the perturbation amplitude normalised by the radius of the fibre, $\zeta/a$, as a function of $t$, for both  no-slip and   slip fibres. The data for both boundary conditions is consistent with the expected exponential growth in the early regime. Thus, the initial slopes of these curves provide reliable measurements of the growth rates.

\begin{figure}[t!]
\includegraphics[width=1\columnwidth]{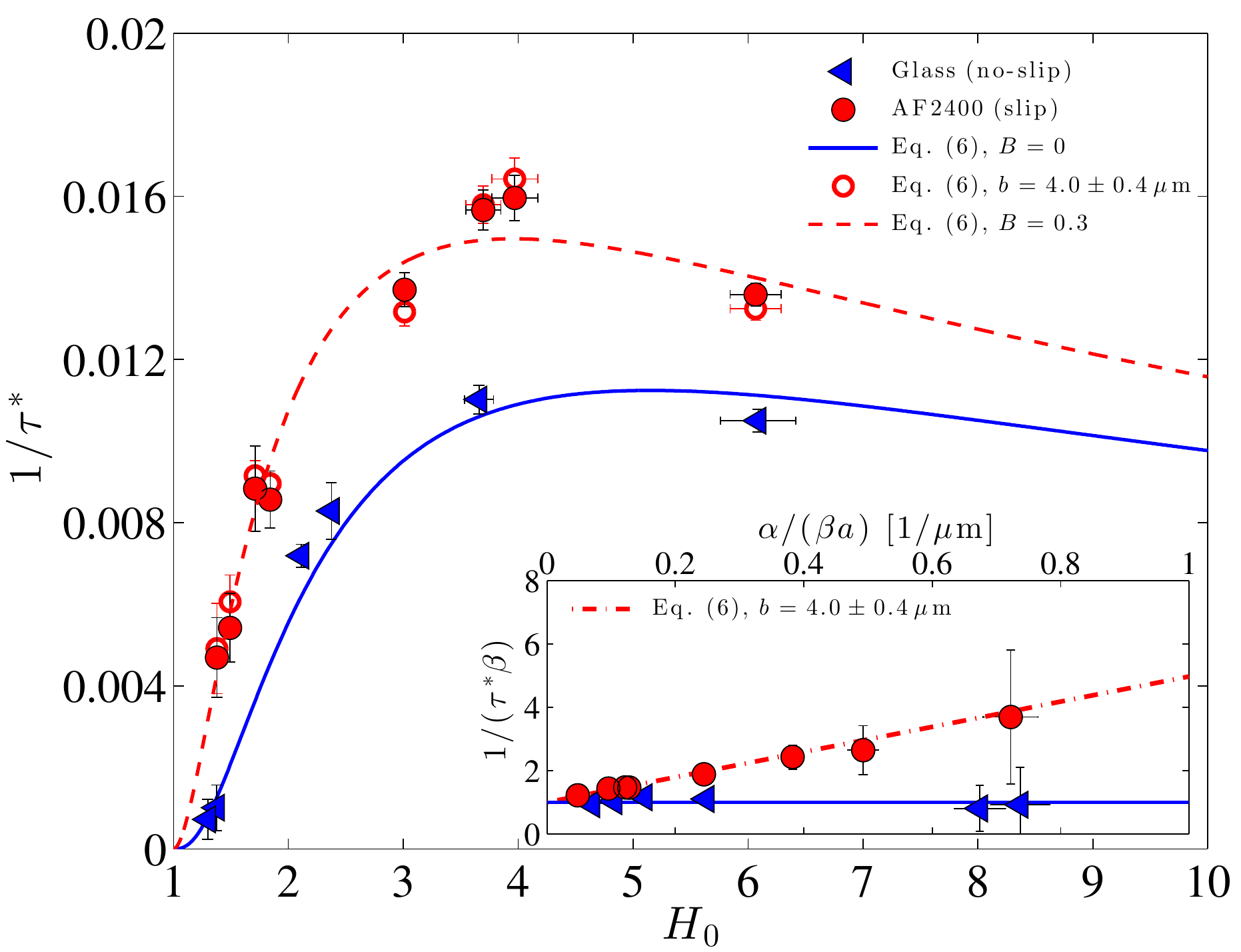}  
\caption{\textbf{Influence of slip on the growth rate.} The inset shows the dimensionless growth rate $1/\tau^*$ normalised by the no-slip case $\beta$, as a function of $\alpha/(\beta a)$,  see Eqs.~(\ref{growth rate}) and~(\ref{growth rate aux}). The slip length $b$ is obtained from a best linear fit (dash-dotted) to the slip data. The error bars are calculated from the error in the geometry and the inaccuracy given by the growth rate measurement. The main curve shows the dimensionless growth rate  of the fastest growing mode on no-slip (glass) and slip (AF2400) fibres, as a function of the dimensionless initial total radius $H_0=1+e_0/a$ (see Fig.~\ref{schematic}~(a)). Open symbols represent growth rates calculated from Eq.~(\ref{growth rate}) using $b=4.0 \pm 0.4\, \mu$m and the respective experimental geometries. Also shown is the theoretical curve for no-slip (Eq.~(\ref{growth rate}), with $B=0$). Furthermore, the theoretical curve for slip (Eq.~(\ref{growth rate}), with $B=0.3$) is plotted as a guide to the eye. }
\label{results}
\end{figure}

The dimensionless growth rates $1/\tau^*$ are shown in Fig.~\ref{results} for both slip and no-slip  fibres, as a function of the dimensionless initial total radius $H_0$. We see that for both the slip and no-slip boundary conditions, the growth rates show a similar geometry dependence. The maxima for the slip and no-slip data can be easily understood: a decreasing growth rate as $H_0$ converges to 1 is due to the diminishing thickness of the liquid film, and thus to the reduced mobility, while the decreasing growth rate for large $H_0$ is due to smaller curvatures and thus a smaller driving force of the instability. 

According to Eq.~(\ref{growth rate}) with $B=0$ (no slip), a maximum in the growth
rate is expected for $H_0 = 1 + e_0/a = 5.15$. This prediction is
entirely consistent with the data. Thus, a coated fibre of a given diameter is maximally unstable when the ratio of the film thickness to the fibre radius is about $e_0/a \sim 4$, consistent with an earlier theoretical study \cite{ROE:1975tw}. The capillary velocity is the only adjustable parameter in the no-slip case.  We obtained $\gamma/\eta=294 \pm 43\,{\mu \text{m}}/{\text{min}}$, which is in agreement with previous data~\cite{McGraw:2013to} adjusted to the temperature used here through the William-Landel-Ferry equation~\cite{MLWilliams:flzZLJWq, Rubinstein:2003vda}. For the slip case, there are now two free parameters: the capillary velocity $\gamma/\eta$ and the dimensionless slip length $B$. If we take the value of the capillary velocity to be that of the no-slip case, we are left with only one true fitting parameter.

To quantify the slip length in our system the growth rates normalised by $\beta$ are plotted as a function of $\alpha/(\beta a)$ (see inset  Fig.~(\ref{results}) and Eqs.~(\ref{growth rate}) and~(\ref{growth rate aux})). As expected, the no-slip data is consistent with the theoretical prediction of $1/\tau^*=\beta$  for all geometries. In contrast  the ratio $1/({\tau^*\beta})$ reveals that the amplification due to slip  on the rise of the instability is more pronounced as $\alpha/(\beta a)$ increases. For the fibre radii used here, large $\alpha/(\beta a)$ corresponds to small $H_0$ (see Eq.~(\ref{growth rate aux})). The stronger influence of slip observed for smaller values of $H_0$ is  due to a non-zero velocity of polymer molecules at the solid-liquid interface ~\cite{Fetzer:2007kt, Baumchen:2009in}. For the smallest value of $H_0$, corresponding to $e_0/a\sim0.4$, the  slip-induced amplification factor of the growth rate is as large as  $\sim 4$. On the contrary, in thick polymer films, the impact of slip is diminished.   From a best linear fit to the slip-data shown in the inset of Fig.~(\ref{results}), the slip length is found to be $b=4.0\pm0.4\, \mu$m. Obtaining a slip length in the range of micrometers is in accordance with former studies on the dewetting of entangled polymer films from substrates with a fluoropolymer coating~\cite{Baumchen:2009in}. 

Having determined the value of $b$ and knowing the fibre radii, we obtain $B$ for each of the experimental geometries. Based on the theoretical model (Eq.~(\ref{growth rate})) corresponding growth rates can be calculated and are shown to be in excellent agreement with the experimental data (see Fig.~(\ref{results})). The dimensionless slip length $B=b/a$ ranges from $0.25$ to $0.37$ in our experiments. To guide the eye, a typical curve calculated with $B=0.3$ is shown in Fig.~(\ref{results}). We see that for the fibres with the slip boundary condition the growth rate is larger than in the no-slip case for a given geometry, and the maximum of the growth rate $1/\tau^*$ corresponds to a smaller value of $H_0$. As expected, a slippery surface facilitates a higher mobility, and hence a faster growth of the instability. This enhanced mass-transport also explains the horizontal shift of the maximum of the growth rate: for a given geometry -- and thus curvature -- the mobility of the slip case is increased in comparison to no-slip. Therefore, the maximum of the growth rate is shifted to lower values of $H_0$. 

\section*{Discussion}
We report on the Plateau-Rayleigh instability of a viscous liquid polystyrene layer on a solid fibre of radius $a$, when the boundary condition at the solid-liquid interface is varied between the classical case of no-slip and the relevant situation of slippage. The wavelength of the fastest growing mode on a slip fibre shows a linear dependence on the initial total polymer-fibre radius, $h_0=a+e_0$, and is not affected by the boundary condition, consistent with the lubrication theory developed. For both slip and no-slip fibres, we observe an exponential temporal growth of the instability at short times, and the respective growth rates show a qualitatively similar geometry dependence. In the case of a slip fibre, the geometry corresponding to the maximum of the growth rate $1/\tau^*$ is shifted to a smaller value of $h_0/a$, and the rise of the instability is faster due to the added mobility at the solid-liquid interface. The slip induced amplification of the growth rate is significant in a parameter range that is of paramount technological relevance. The linear stability analysis of the thin film equation developed here is in excellent agreement with the data, valid for all Newtonian liquids, and provides a robust measure of two fundamental quantities: the capillary velocity $\gamma/ \eta$, and the slip length~$b$.

\section*{Methods}
\subsection*{Preparation of fibres}
Glass fibres were prepared by pulling heated glass capillary tubes to final radius in the range $10 < a <25 \ \mu$m  using a pipette puller (Narishige, PN30).  To prepare the slip boundary condition, the glass fibres were hydrophobised by dip-coating in a 0.5 wt\% solution of AF2400 (Poly[4,5-difluoro-2,2-bis(trifluoromethyl)-1,3-dioxole-co-tetrafluoroethylene]) (Aldrich) in a perfluoro-compound solvent (FC72\textsuperscript{TM}, Fisher Scientific), to form an amorphous fluoropolymer layer ($T_\textnormal{g} \sim 240\,^\circ$C). A dip-coating speed of 1$\,$mm/s resulted in an AF2400 layer with a thickness of $14 \pm 1\,$nm. The hydrophobised fibres were annealed in a vacuum chamber at $80\,^\circ$C for $90$ minutes to remove excess solvent. 
\newline

{\subsection*{Preparation of homogeneous PS films}
To prepare homogeneous PS films, a concentrated solution (35 wt\%) of atactic polystyrene (Polymer Source Inc.) with a molecular weight of $78\,\mathrm{kg/mol}$ and low polydispersity ($M_{\textrm{w}}/M_{\textrm{n}}=1.05$) was dissolved in chloroform (Fisher Scientific). A droplet of the highly viscous polymer solution was placed between two glass slides, forming a meniscus at the edge of the glass slides. We note that the chloroform does not dissolve the underlying AF2400 coating on the slip fibres. A slip or no-slip fibre could then be placed in the middle of the gap between the slides and pulled out of the droplet with a constant speed using a motorised linear translation stage. By varying the pulling speed $v_0$ in the range $80<v_0<150\,$mm/s, we obtained film thicknesses that ranged from $e_0=5$ to $93\,\mu$m after the solvent had evaporated. 
\newline

\subsection*{Experimental setup}
The as-prepared samples were placed into a heated sample cell to initiate the PRI. With two $\sim 0.5$~mm thick spacers, the coated fibres were suspended above a reflective Si wafer (to improve contrast), and placed on a microscope hot stage (Linkam). A metal ring in direct contact with the hot stage supported a glass cover over the sample and Si wafer ({see Supplementary Figure 1 and Supplementary Methods}), thereby ensuring good thermal contact and temperature control to within 1$^\circ \mathrm{C}$. The surface profiles were analysed from the optical micrographs taken at various times using a custom-made edge detection software written in MATLAB. 

\section*{Acknowledgements}The authors wish to thank Martin Brinkmann for insightful discussions.
The authors thank  the graduate school GRK 1276, NSERC of Canada, the German Research Foundation (DFG) under grant numbers BA3406/2 and SFB 1027 for financial support.

\end{document}

\section*{Supplementary Figures}

\begin{figure}[h]
\begin{center}
\includegraphics[width=0.6\columnwidth]{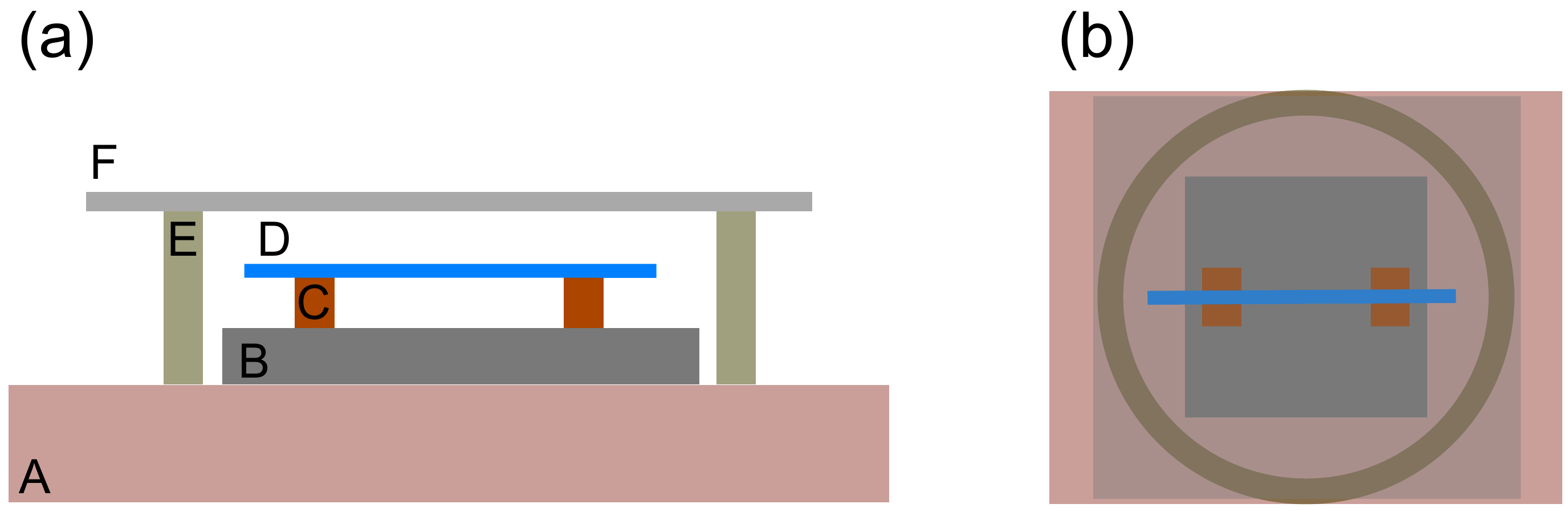}  
\caption{\textbf{Schematic of the sample chamber.} Representation of (a) the side view and (b) the top view of the experimental setup. On top of a hot stage (A), a Si wafer (B) with two teflon spacers (C) is placed. Freely suspended filaments are realized by depositing the fibre (D) across the spacers. The fibre is surrounded by a metal ring (E) with a glass cover (F).}
\label{setup}
\end{center}
\end{figure} 

\newpage

\section*{Supplementary Method}

\bigskip
\noindent \textbf{Fibre thin film equation with slip:}
\bigskip

Here, we provide the main steps of the derivation of the fibre thin film equation (see Eqs.~(2) and (3) in the article). Let us consider a viscous liquid film coating a cylindrical fibre of constant radius $a$  (see Fig.~1 in the article). The axis of the fibre, along the $x$ direction, is chosen as a reference for the radial coordinate $r$. We assume cylindrical invariance of the free-surface profile described by $r=h(x,t)$, at position $x$ and time $t$. Furthermore, we assume an incompressible flow of the viscous Newtonian fluid, where gravity, disjoining pressure and inertia are negligible. Defining the local excess pressure $p(r,x,t)$ and the velocity field $\boldsymbol v(r,x,t)$, and assuming the lubrication approximation to first order, namely that  the slopes of the profiles remain small and the velocity is mainly oriented along the fibre axis, $\boldsymbol v\,=\,v\boldsymbol e_{x}$, leads to the two projections of the Stokes equation:
\begin{subeqnarray}
\label{condsup}
\slabel{condsup1}\partial_xp&=&  {\eta}\left(\partial_{rr}v +\frac{\partial_rv}r \right)\\
\slabel{condsup2} \partial_r p&=&0 \ ,
\end{subeqnarray}
where $\eta$ is the dynamical shear viscosity that we assume to be homogeneous and constant. The lubrication approximation is valid since the profile slopes remain small in comparison to $1$ at early times -- the regime in which the linear analysis is performed -- and since the typical horizontal length scale of the flow is expected to be larger than the minimal wavelength of the Plateau-Rayleigh instability, that is given by $\lambda_{\textrm{min}}=2\pi h_0\gg e_0$. This is indeed always the case as even in the worst case geometry, $e_0 \approx 5a$, one has $e_0/\lambda_{\min} \approx0.13\ll1$. 
According to {Supplementary} Eq.~\eqref{condsup2}, the excess pressure is invariant in the radial direction, to first order in the lubrication approximation. Thus, the excess pressure is set by the Laplace boundary condition at the free surface which, within the small slope approximation, reads:
\begin{eqnarray}
\label{Laplacesup}
p&=&\gamma\left(\frac1h-h''\right) \ ,
\end{eqnarray}
where $\gamma$ denotes the air-liquid surface tension, taken to be homogeneous and constant, and where the prime denotes the derivative with respect to $x$. The two terms on the right hand side of {Supplementary} Eq.~\eqref{Laplacesup} correspond to the radial and axial curvatures of the free surface. Note that, as already mentioned in the body of the article, we have kept a second order lubrication term in the pressure contribution: the axial curvature $-h''$. This is because it is the lowest order term counterbalancing the driving radial curvature $1/h$, and it is thus crucial to obtain the actual threshold of the instability [9]. Regarding the boundary conditions, we assume no shear at the free surface and the Navier slip condition at the solid-liquid interface, namely:
\begin{subeqnarray}\label{BCsup}
\slabel{BCsup1}\partial_rv|_{r=h}&=&  0\\
\slabel{BCsup2} \partial_r v|_{r=a}&=&\frac{v|_{r=a}}b \ ,
\end{subeqnarray}
where $b$ denotes the slip length. Integrating {Supplementary} Eq.~\eqref{condsup}, together with {Supplementary} Eqs.~\eqref{Laplacesup} and~\eqref{BCsup}, yields:
\begin{eqnarray}
\label{Vitessesup}
v&=&\frac{\gamma\left( h'+h^2h'''  \right) }{4\eta h^2}\left[ 2h^2\log\left(\frac ra\right)  -r^2 +\frac{ 2b}a h^2 +a^2-2ab \right] \ ,
\end{eqnarray}
for all $r\in [a,h]$. Volume conservation requires that:
\begin{eqnarray}
\label{VolConssup}
\partial_th +\frac{Q'}h&=&0 \ ,
\end{eqnarray}
where we introduced the volume flux per radian:
\begin{eqnarray}
Q&=&\int_{a}^h\text d r\, r v \ . \label{Debitsup}
\end{eqnarray}
Combining {Supplementary} Eq.~\eqref{VolConssup} together with {Supplementary} Eqs.~\eqref{Vitessesup} and \eqref{Debitsup} yields the fibre thin film equation:
\begin{eqnarray}
\label{FTFEdim}
\partial_th +  \frac{\gamma}{16 \eta h}\left[ \left(h'+h^2h''' \right)\left( 4h^2\log \left(\frac ha\right)  +\left(\frac{4b}{a}-3\right)h^2+4a^2-8ab+\left(4ab-a^2\right)\frac{a^2}{h^2}  \right) \right]' =\,\,0 \ .
\end{eqnarray}
Finally, introducing the dimensionless variables (see Eq.~(1) in the article) into {Supplementary} Eq.~\eqref{FTFEdim} yields the dimensionless fibre thin film equation (see Eqs.~(2) and (3) in the article).

\bigskip
\noindent \textbf{Experimental setup and details:}
\bigskip

The as-prepared samples were annealed in ambient atmosphere at $180\,^\circ$C - well above $T_{\textnormal{g}}$ - which causes the polystyrene films to melt and the PRI to develop. The evolution of the PRI  was recorded with an inverted optical microscope (Olympus BX51). At various times, images were taken using a 5x magnification objective and a camera with a resolution of 1392 x 1040 pixels (QImaging, QIClick), resulting in a pixel size of $1.28\,\mu$m. To ensure free standing filaments, the coated fibres (D) were placed along two $\sim 0.5\,$mm thick teflon spacers (C) located on a Si wafer (B) (see Supplementary Figure \ref{setup}).  The Si wafer was used to improve the optical contrast. A metal ring (E) in direct contact with the hot stage supported a glass cover (F) over the sample and Si wafer. The temperature within the sample cell is uniform (within $1^{\circ}$C) and convection does not affect the measurements. By using spacers (C) made of teflon, we ensured that there is no drainage of the liquid PS film towards the edges (PS does not wet teflon). The surface profiles were analysed from the optical micrographs taken at various times, using a custom-made edge detection software written in MATLAB.